\documentclass[aps,pra,showpacs]{revtex4}
\usepackage{graphicx}
\usepackage{comment}
\begin{document}

\title{Frustrated total internal reflection and the illusion of
superluminal propagation}

\author{Vera L. Brudny}
\email{vera@df.uba.ar}
\affiliation{Departamento de F\'{\i}sica, Facultad de Ciencias Exactas
y Naturales, Universidad de Buenos Aires, Cd. Universitaria,
Pabell\'on 1, C1428 EHA Buenos Aires, Argentina}
\author{W. Luis Moch\'an}
\email{mochan@fis.unam.mx}
\affiliation{Instituto de Ciencias F\'{\i}sicas, Universidad Nacional
  Aut\'onoma de M\'exico, Apartado Postal 48-3, 62251 Cuernavaca,
  Morelos, M\'exico} 
\date{\today}

\begin{abstract}

We analyze the propagation of a pulse across a vacuum gap separating 
opposite flat parallel faces of two
transparent dielectrics by means of an explicitly causal and
retarded propagator constructed directly from the free-space wave
equation. Nevertheless, our approach yields apparently superluminal
propagation for the case of frustrated total internal reflection
(FTIR), that is, a transmitted wave packet appears on the far side of
the gap at the same time that the corresponding incident packet
crosses the front one.  Thus, in this example superluminality is just
an illusion, being consistent with both casuality and classical
electrodynamics.  We study the origin of the apparent superluminality
in this case, which is inherent to light pulse propagation in free
space and does not depend on the particulars of light-matter
interaction, and find that it is due to propagation from the lateral
wings of the incident pulse to the central part of the
transmitted pulse. Thus, notwithstanding their similarities, FTIR is
not equivalent to 1D tunneling. We propose experiments to test our
explanation of superluminality using opaque screens to block part of
the wavefront, although we demonstrate that the
propagation of smooth finite pulses constrained to be made up
completely of evanescent Fourier components is
indistinguishable from truly superluminal propagation, i.e., it may be
completely accounted for using an explicitely superluminal and acausal
propagator as well as the causal subluminal one.

\end{abstract}

\pacs{
42.25.Bs, 
41.20.Jb, 
03.65.Xp, 
42.25.Gy 
}

\maketitle

\section{Introduction}

The physics of light propagation is a topic of active research, due to
its relevance to several technological applications and basic
research. Recent research on new materials has shown that it is
possible to exercise an extraordinary control on the propagation of
light pulses, which in turn has generated new interest in both
practical and fundamental questions on light propagation. Moreover,
despite the fact that Maxwell completed the formulation of the
classical theory of electromagnetism in 1864 and Einstein's special
theory of relativity was presented in 1905, some hot controversies
continue to arise on the subject of superluminal propagation
\cite{wang07,winful07,pereyra,ranfagni07,winful05,winful03,buttiker03,winful03-1,buttiker03-1,winful03-2}
and its possible implications for both classical and quantum
information theory \cite{qi,de_angelis}.  Recent reviews on the
subject can be found in Refs. \cite{winful06,boyd02}

One consequence of the special theory of relativity is that no signal
can cause an effect outside the light cone of its source.  Violation of this
principle of relativistic causality leads to paradoxes such as that of
an effect preceding its cause \cite{azbel,garrison1998}.  When dealing
with light propagation in a material characterized by a given dispersion
relation $\omega(\vec k)$ between the frequency $\omega$ and the wave vector
$\vec k$, several {\em velocities} may be
defined, such as the {\em phase velocity} $v_\phi = \omega/k$ and the {\em
  group velocity} $v_g=\nabla_{\vec k}\omega$. It is
recognized that under certain conditions \cite{jackson} both
velocities can exceed the speed of light in 
vacuum $c$. This does not
contradict the postulates of the special theory of relativity, for it
has been recognized since the works of Sommerfeld \cite{sommerfeld}
and Brillouin \cite{brillouin-wp} that in order not to violate the
principle of causality it is the {\em information velocity} that must
not exceed $c$.  The question that arises then is what is an
appropriate and operative definition of {\em information
  velocity}. There have been several discussions and proposals on this
subject, but the question is not yet settled
\cite{diener96,diener97,kuzmich,stenner,stenner05,ranfagni06}.

The vast majority of the published work concerning superluminal pulse
propagation deals with light propagation in material media, and the
usual analysis attempts to explain how the interaction of
electromagnetic radiation with the medium affects the propagation of
light pulses in such a way that they appear to travel
superluminally.  In a pioneering paper \cite{icsevgi}, Icsevgi and
Lamb performed a theoretical investigation of the propagation of
intense laser pulses through a laser amplifier. Apparent superluminal
light propagation has been reported in gain-assisted systems
\cite{wang00,janowicz,huang} as well as in birefringent crystals
\cite{solli03,brunner,halvorsen}, composite media and photonic
crystals \cite{kulkarni,safian} and dispersive media
\cite{bigelow,talukder}.

There have also been claims of evidence of superluminal propagation in
free space \cite{mugnai00,mugnai05} and during {\em optical tunneling} in
frustrated total internal reflection (FTIR) configurations
\cite{carey00,mochan01,carey01,shaarawi}. Optical tunneling has been
studied by several authors, both theoretically and experimentally
\cite{oe,barbero,reiten,balcou97,resch}. Some of the observed results
are still subject of debatable interpretation \cite{winful06} and do
not close the subject of whether there are possibilities for
superluminal transmission of information in such systems. We have
therefore chosen to address this subject in way that leads to
straightforward interpretation of the results while resorting only to
classical electromagnetic theory. We claim that although some results
may {\em appear} to indicate superluminal propagation, there is no
real superluminal transfer of information.

In this paper we analyze mathematically the
propagation of a pulse across a vacuum gap separating opposite flat
parallel faces of two transparent
dielectrics by means of an explicitly causal and
retarded propagator constructed directly from the free-space wave
equation. Our results yield indeed an apparent
superluminal propagation corresponding to the conditions of FTIR,
but they show explicitly that it is consistent with both casuality and
with classical electrodynamics. Our example shows
superluminality effects inherent to light pulse propagation in free
space which therefore does
not depend on the particulars of light-matter interaction.  The {\em
  illusion} of 
superluminality consists of transmitted pulses arriving to the
far side of the gap in synchrony with the crossing of the front
surface by the incident pulse.  We explain this illusion of
superluminal behavior in terms of a causal, subluminal propagation, 
taking into account the spatial extent of the incident pulse along its
transverse as well as its longitudinal directions and we propose
experiments to demonstrate the retarded nature of propagation in
FTIR. Nevertheless, we find that for constrained pulses fully made up of
evanescent Fourier components, subluminal and superluminal
propagation in FTIR experiments are indistinguishable.

This paper is organized as follows. We first introduce a propagator
that will allow the analyzis of electromagnetic pulse propagation
across a vacuum gap
(Section \ref{secpropagator}). This propagator is both casual and
retarded and complies with the classical electromagnetic theory.  We
then study the propagation across the gap of pulses that arrive as
plane waves with well defined angles of incidence. In Section
\ref{secnormal} we study the case of
subcritical angles, yielding non-evanescent transmitted waves. In
Section \ref{secevan} we study the case of hypercritical angles,
yielding evanescent waves. We conclude that the propagation of a
light pulse in a FTIR configuration may appear superluminal and
acausal but that it is actually subluminal and that propagation
has to account necessarily for the lateral wings of the incident
pulse. In Section \ref{secscreens} we suggest  experiments that
might demonstrate  the actual causal and subluminal nature of the
apparent superluminal behavior by using sharp opaque screens that
block parts of the incident wave so that its extent becomes finite
along both its propagation and its transverse
directions. Nevertheless, as the borders of these screens  
produce propagating diffracted waves,
in Section \ref{secsmooth} we eliminate them and we study incident
pulses that are finite along several spatial
directions but that have a smooth profile. We obtain that if they are 
comprised of evanescent Fourier components only, they appear
to propagate 
superluminally through the vacuum gap, even though their behavior is determined
by our causal retarded propagator. In Section \ref{secevprop} we construct an
alternative acausal, superluminal propagator, and prove that it is exactly
equivalent to the causal and retarded propagator when applied to fully
evanescent finite pulses. Thus, for such constrained pulses, it is
impossible to distinguish causal subluminal from acausal and
superluminal propagation;  the illusion of superluminality
appears to be 
not only a matter of interpretation of the {\em result} of the
propagation, but may be also present in the description of the
propagation process itself. We present our conclusions in Section
\ref{secconclusion}.

\section{Propagator}\label{secpropagator}
\begin{figure}
\includegraphics{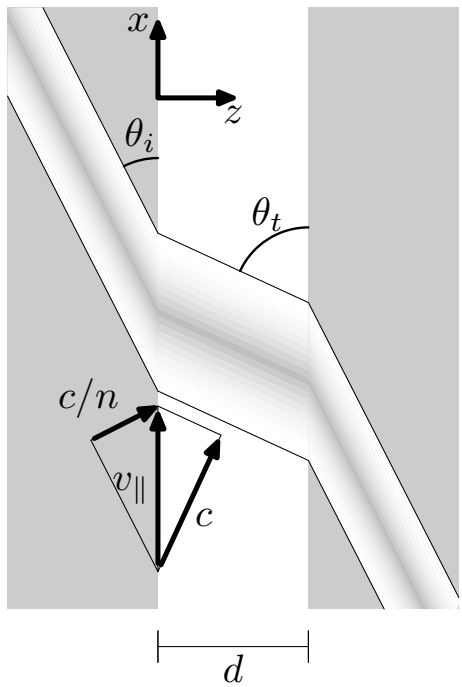}
\caption{\label{normal}Pulse with arbitrary profile incident at an
angle $\theta_i<\theta_c$ onto the surface 
$z=0$ of a dielectric with index of refraction $n=2$. The pulse is
transmitted into a vacuum gap and into a second dielectric at
$z=d$. The speed of propagation within each media, $c/n$ and $c$, are
indicated, as well as the speed of propagation of the wavefronts along
the interfaces $v_\parallel$.} 
\end{figure}

Consider two transparent dielectrics occupying the regions $z\le0$ and
$z\ge d$. In this section we obtain the causal and retarded propagator
that 
describes the motion of a pulse across a vacuum gap $0<z<d$ spanning from the
interface at $z=0$ to that at $z=d$ (Fig. \ref{normal}).
As we want to consider explicitly the angle of incidence onto the
interface $z=0$, we cannot treat the problem beforehand as if it were 1D. 
For simplicity, we will assume full translational symmetry along the
$y$ direction, so that our problem becomes 2D.
Thus, we start with the scalar wave equation
\begin{equation}\label{greenseq}
(\nabla^2-\frac{1}{c^2}\frac{\partial^2}{\partial
t^2})G_0(t,x,z;t',x',z') =  \delta(t-t') \delta(x-x') \delta(z-z')
\end{equation}
with a unit singular point source fired
at time $t'$ at position $(x',z')$, which is solved by the
explicitly causal 
and retarded free space Green's function 
\cite{inmorse}
\begin{equation}\label{G0}
G_0(t,x,z;t',x',z')=\frac{c}{2\pi}\frac{\Theta\left(c(t-t')-\sqrt{(x-x')^2+(z-z')^2}\right)}{\sqrt{c^2(t-t')^2-(x-x')^2-(z-z')^2}},
\end{equation}
where $\Theta(\ldots)$ is the Heaviside unit step function. Using
image theory we can construct a Green's function that obeys 
Dirichlet boundary conditions at the surface $z=0$,
\begin{equation}\label{G}
G(t,x,z;t',x',z')=G_0(t,x,z;t',x',z')-G_0(t,x,z;t',x',-z').
\end{equation}
In the half-space $z>0$ Green's theorem yields the solution
\cite{inmorse}
\begin{equation}\label{propagate}
\phi(t,x,z)=\int dx' \int dt'\,P(t,x,z;t',x',0^+) \phi(t',x',0^0)
\end{equation}
of the homogeneous scalar wave equation 
that is outgoing as $z\to\infty$ and is null in the remote past, where
$\phi(t',x',0^+)$ denotes its previous values on the boundary $z=0^+$,
and 
\begin{equation}\label{propagator}
P(t,x,z;t',x',0^+)=\frac{\partial}{\partial
z'}\left. G(t,x,z;t',x',z')\right|_{z'=0^+}
\end{equation}
is the causal retarded propagator of the problem. We may identify the
field $\phi$ with the component $E_y$ of the electric field $\vec E$
in the case of a TE or $s$ polarized incoming wave, and with the
component $B_y$ of the magnetic field $\vec B$ in the case of TM or
$p$ polarization. The propagator
(\ref{propagator}) does not account for the presence of the two
dielectrics bounding the air gap. Thus, the field (\ref{propagate})
has no
information about the multiple reflections at the boundary of the
gap. In principle, 
these can be incorporated by reflecting the field at
the interfaces $z=0,d$ using the appropriate Fresnel coefficient
and propagating it back and forth across the gap with the propagator
(\ref{propagator}) for the $z=0$ surface and a similar one for the
$z=d$ surface. The total field would then be the sum of all the
multiply reflected fields and would have information about the
electromagnetic properties of the reflecting surfaces. In this paper
we will restrict ourselves to an analysis of the first crossing of the
air gap $0^+\to d^-$, and thus our results will be unrelated to the
nature of the bounding media.

Substituting 
Eq. (\ref{G0}) into (\ref{G}) and (\ref{propagator}) we obtain
\begin{equation}\label{P}
P(t,x,z;t',x',0^+)=-\frac{c}{\pi} \frac{\partial}{\partial z} 
\frac{\Theta\left(c(t-t')-\sqrt{(x-x')^2+z^2}\right)}{\sqrt{c^2(t-t')^2-(x-x')^2-z^2}}.
\end{equation}
The field $\phi$ can then be written in terms of an ancillary
function 
\begin{equation}\label{phivspsi}
\phi(t,x,z)=-\frac{\partial}{\partial z}\psi(t,x,z),
\end{equation}
where
\begin{equation}\label{psi}
\psi(t,x,z) =\frac{c}{\pi} \int dx' \int dt'\,
\frac{\phi(t',x',0^+)}{\sqrt{c^2(t-t')^2-(x-x')^2-z^2}},
\end{equation}
plays the role of a potential and the integration is performed within
the region 
$c(t-t')>\sqrt{(x-x')^2+z^2}$. Clearly, the procedure above yields a
causal (sub)luminal propagation from the $z=0^+$ plane to any point in
the $z>0$ vacuum.

\section{Non-Evanescent wave transmission}\label{secnormal}

We consider now that an arbitrarily shaped pulse impinges {\em at a
  well defined angle} $\theta_i$ on the inside surface $z=0^-$ of a
homogeneous non-dispersive dielectric with index of refraction $n$
(Fig. \ref{normal}).  The incident pulse is therefore described by an
arbitrary function $\phi_i(t,x,z)=f_i[t-(n/c)\hat n_i\cdot\vec\rho]$
of a single variable $t-(n/c)\hat n_i\cdot\vec\rho$, where $\hat
n_i\equiv (\sin\theta_i,\cos\theta_i)$ is a unit vector pointing along
the angle of incidence $\theta_i$ and $\vec \rho\equiv(x,z)$. Notice
that in this case the incident wavefronts have an infinite extension
in the direction normal to $\hat n_i$. At $z=0^+$ and after being
transmitted into vacuum, the outgoing field can therefore be written
as
\begin{equation}\label{phivsf}
\phi(t,x,0^+)=f_t(t-x/v_\parallel)
\end{equation}
where $f_t$ is related to the arbitrary function $f_i$ and the Fresnel
amplitude for transmission from the dielectric into vacuum. As we are
concerned only with propagation across the vacuum gap, we will take $f_t$
as given and we will disregard its relation with $f_i$, which would
involve the dielectric properties of the incident medium.

The
intersection of the pulse with the interface $z=0$ 
is therefore seen to travel along $x$ with velocity
$v_\parallel=c/(n\sin\theta_i)$ (not to be confused with the parallel
component of the incident velocity $\hat n_i c/n$). For incidence angles 
smaller than the critical angle $\theta<\theta_c=\sin^{-1}(1/n)$,
$v_\parallel>c$ and we have normal transmission, while for
$\theta>\theta_c$, $v_\parallel<c$ and 
total internal reflection ensues.

In order to set up a reference with which to compare the evanescent
case, in this section we employ our propagator to 
study a non-evanescent plane pulse. Thus,
we consider here the case $\theta<\theta_c$ and we substitute
Eq. (\ref{phivsf}) 
into Eq. (\ref{psi}) to obtain
\begin{equation}\label{psinormal}
\psi(t,x,z)=\frac{c}{\pi} \int dt''\, f_t(t'') \int dx'\,
\left(c^2(t-t''-x'/v_\parallel)^2 - (x-x')^2-z^2\right)^{-1/2}
\end{equation}
after changing integration variables from $t'$ to $t''\equiv
t'-x'/v_\parallel$. The integration (\ref{psinormal}) has to be
performed over the region $c(t-t'')-\mu x'>\sqrt{(x-x')^2+z^2}$, where
$\mu\equiv c/v_\parallel$. Thus, $t''$ has an upper bound
\begin{equation}\label{tmax}
t_m=t-(\mu x - \nu z)/c,
\end{equation}
where $\nu\equiv\sqrt{1-c^2/v_\parallel^2}$, 
and for each value of $t''<t_m$, $x'$ is bounded by the limits 
\begin{equation}\label{xmasmenos}
x'_\pm=-(1/\nu^2)\left(x-\mu c(t-t'') \pm\sqrt{[\mu
    x-c(t-t'')]^2-\nu^2 z^2}\right).
\end{equation} 

\begin{figure}
\includegraphics{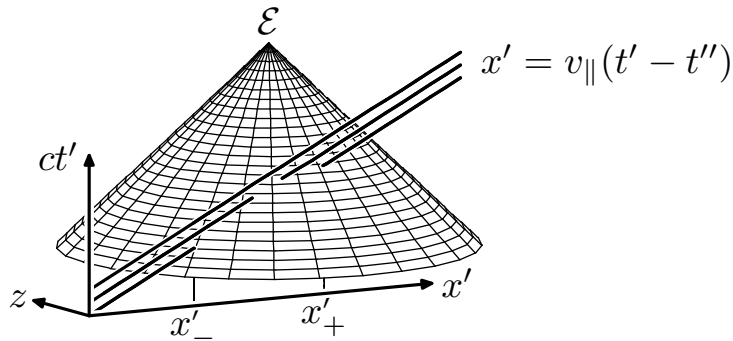}
\caption{\label{cono1}Contributions to the non-evanescent wave observed
at an event ${\cal E}=(t,x,z)$. Some world lines
$(ct',x'=v_\parallel(t'-t''),0)$  of source points labeled by
fixed values of $t''$ and thus moving on the $z=0$ plane along
with the incident pulse are shown. The positions $x'_\pm$ denote the
intersections of the past light cone of $\cal E$ with one of the world
lines plotted. We consider the case $v_\parallel>c$. The
coordinate and time axes are indicated.} 
\end{figure}

A simple interpretation of Eqs. (\ref{tmax}) and (\ref{xmasmenos}) can
be obtained with the help of Fig. \ref{cono1}.
Consider an event ${\cal E}=(ct,x,z)$ defined by the observation of
the field at a given position $(x,z)$ with $z>0$ and at a given time
$t$. Causality requires that only events within the past light-cone of
$\cal E$ are able to influence it.  Notice that a given value of $t''$
denotes a point that moves along $x$ keeping a fixed position with
respect to the
intersection of the incident pulse with the $z=0$ interface. In
Fig. \ref{cono1} we show the world lines $(ct',x',0)$ of a few such
points. Since each of them moves with speed $v_\parallel>c$, most of
its world line lies outside the past light cone of $\cal E$. Only if
$t''<t_m$ can it actually cross the light cone, entering and leaving
at positions $x'_-$ and $x'_+$ respectively.

After the change of variables from $x'$ to
\begin{equation}\label{eta}
\eta\equiv \frac{x'-[x-\mu c(t-t'')]}{\sqrt{[\mu
x-c(t-t'')]^2-\nu^2 z^2}}, 
\end{equation}
Eq. (\ref{psinormal}) simplifies to
\begin{equation}\label{psivseta}
\psi(t,x,z)=\frac{c}{\pi \nu} \int_{-\infty}^{t_m} dt''\,f_t(t'')\int_{-1}^1\frac{d\eta}{\sqrt{1-\eta^2}}.
\end{equation}
The integration over $\eta$ is immediate, so that substituting
Eq. (\ref{psivseta}  in Eq.(\ref{phivspsi}) we obtain finally
\begin{equation}
\phi(t,x,z)=f_t(t-\hat n_t\cdot\vec\rho/c).
\end{equation}
where $\hat n_t=(\mu,\nu)$.  Thus, as illustrated in
Fig. \ref{normal}, the transmitted wave is a pulse with the same
profile as the incident field and propagating with speed $c$ at the
well defined angle $\theta_t=\sin^{-1}\mu=\cos^{-1}\nu$, in accordance
with Snell's law as could have been expected.

\section{Evanescent wave transmission}\label{secevan}

\begin{figure}
\includegraphics{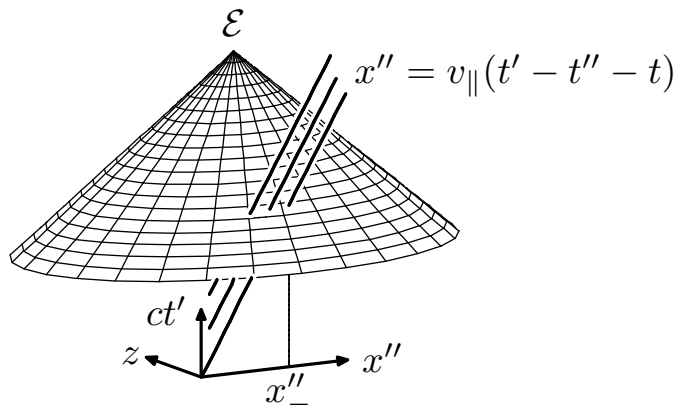}
\caption{\label{cono2}Contributions to the evanescent wave observed at
  an event ${\cal E}=(ct,x,z)$ as in Fig. \ref{cono1}, but for
  $v_\parallel<c$. The position $x''_-$ denotes the intersections of
  the past light cone of $\cal E$ with one of the world lines
  $x''=v_\parallel(t'-t''-t)$ corresponding to $t''$.}
\end{figure}

We consider now the case $\theta_i>\theta_c$, for which
$v_\parallel<c$ and the transmitted wave becomes evanescent. In this
case we substitute 
Eq. (\ref{phivsf}) into Eq. (\ref{psi}) to obtain
\begin{equation}\label{psievan}
\psi(t,x,z)=\frac{c}{\pi}\int dt'' f_t(t-x/v_\parallel+t'')\int dx''
\left(c^2(t''+x''/v_\parallel)^2 - (x'')^2-z^2\right)^{-1/2},
\end{equation}
after introducing the variables $x''\equiv x'-x$ and
$t''=t'-t-x''/v_\parallel$. As shown in Fig. \ref{cono2}, for any
observation event $\cal E$ and any value of $t''$, there is exactly
one intersection $x''_-$ between the  past light cone of $\cal E$ and
the world line $(ct',x''=v_\parallel(t'-t''-t),0)$, where now
$x''_-=-\gamma \beta [\gamma c t'' + \sqrt{z^2+(\gamma\beta c
t'')^2}]$ and we introduced the definitions $\beta\equiv
v_\parallel/c$ and $\gamma\equiv1/\sqrt{1-v_\parallel^2/c^2}$. We have
assumed that $v_\parallel>0$. 
As the world
line leaves the past light cone at $x''_-$, in
Eq. (\ref{psievan}) the integration over $t''$ 
is unconstrained and that over $x''$ extends from $-\infty$ to $x''_-$.
Another change of variable, from $x''$ to 
\begin{equation}\label{etaagain}
\eta=\frac{1}{\gamma\beta}\frac{x''+\gamma^2\beta c
t''}{\sqrt{z^2+(\gamma\beta c t'')^2}}, 
\end{equation}
yields
\begin{equation}\label{psianew}
\psi(t,x,z)=\frac{\gamma\beta c}{\pi}\int_{-\infty}^\infty dt''\,
f_t(t-x/v_\parallel+t'') \int_{-\infty}^{-1}
\frac{d\eta}{\sqrt{\eta^2-1}}.
\end{equation}

The integrals over $x''$ and over $\eta$ in 
Eqs. (\ref{psievan}) and (\ref{psianew}) respectively yield an
infinite value for 
$\psi$. This divergence is not unlike that commonly found for the
electromagnetic potentials produced by infinitely  extended
sources. For example, when calculating the electric field 
produced by a uniformly charged plane one cannot simply obtain the
corresponding potential by integrating the Coulomb kernel over the whole
surface. However, in that case the electric field may be obtained
either by deriving the Coulomb kernel first and integrating afterwards
or else, by
truncating the integrations at a finite distance, deriving the resulting
potential to obtain the field and 
afterwards taking the limit of an infinite  
surface.
Here we follow the later procedure.
Thus, we set a finite lower integration limit
$x''_L$ in Eq. (\ref{psievan}), corresponding to a lowest point
$\eta_L$  in Eq. (\ref{psianew}), and we take the limit
$x''_L\to-\infty$, $\eta_L\to-\infty$ after obtaining the field $\phi$.

As $\psi$ depends on $z$ only through $\eta_L$, substituting
Eq. (\ref{psianew}) in (\ref{phivspsi}) we obtain
\begin{equation}\label{phivsetaL}
\phi(t,x,z)=\frac{\gamma\beta c}{\pi}\int_{-\infty}^\infty dt''\,
f_t(t-x/v_\parallel+t'')
\frac{1}{\sqrt{\eta_L^2-1}}\frac{\partial\eta_L}{\partial z}. 
\end{equation}
In the limit $x''_L\to-\infty$ we evaluate 
\begin{equation}\label{detadz}
\xi_L \equiv \frac{1}{\sqrt{\eta_L^2-1}}\frac{\partial\eta_L}{\partial
z}\to  \frac{z}{z^2+(\gamma\beta c t'')^2},
\end{equation}
and we obtain finally
\begin{equation}\label{phievan}
\phi(t,x,z)=\frac{1}{\pi}\int_{-\infty}^\infty dt''\,
f_t(t-x/v_\parallel+t'') \frac{\gamma |v_\parallel| z}{z^2+(\gamma
  v_\parallel t'')^2}.
\end{equation}
Taking the absolute value of $v_\parallel$ in the numerator of
Eq. (\ref{phievan}) allows its use also for $v_\parallel<0$. 

\begin{figure}
\includegraphics{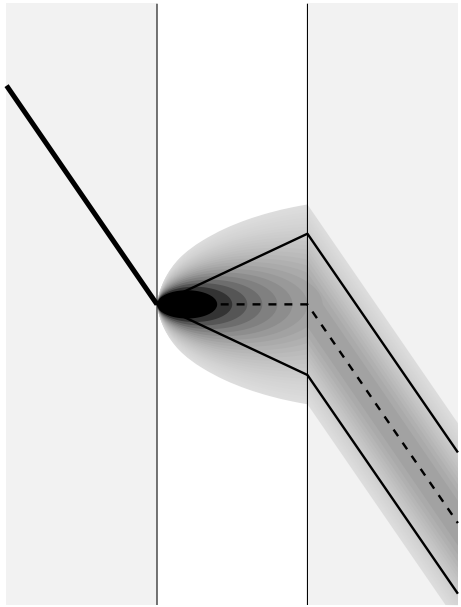}
\caption{\label{figlorentz}Infinitely sharp wavefront (heavy solid
line) incident at an angle $\theta=34.5^\circ>\theta_c$ upon the
surface of a dielectric with index of refraction $n=2$. The pulse
widens and diminishes as it is transmitted across an air gap and into a
second dielectric. The thin lines indicate the nominal pulse width and
the dashed line its center. 
}
\end{figure}

To grasp the meaning of Eq. (\ref{phievan}) we evaluate it for an
infinitely sharp pulse
\begin{equation}\label{sharp}
f_t(\tau)\equiv f_0 \delta(\tau).
\end{equation}
Substitution into Eq. (\ref{phievan}) yields
\begin{equation}\label{lorentz}
\phi(t,x,z)=\frac{f_0}{\pi}  \frac{\gamma |v_\parallel| z}{z^2+
\gamma^2(x-v_\parallel t)^2}.
\end{equation}
Surprisingly,  at any time $t$ the pulse transmitted at a distance $z$
from the 
interface is given by a Lorentzian of width
$z/\gamma$ centered in front of the actual position
$x=v_\parallel t$ of the incident pulse on the $z=0$ surface
(Fig. \ref{figlorentz}).
Thus, the propagation seems to be instantaneous in the direction
normal to the surface, and actually, part of the pulse seems to travel
backwards in time \cite{carey00,resch}, as at a position $(x,z)$ it
becomes appreciable at times $t<x/v_\parallel$, that is, before the
incoming pulse reaches the corresponding position $(x,0)$. However,
our deduction of Eq. (\ref{sharp}) shows that it is completely
consistent with a causal and retarded propagation, and that the field
at a $(x,z)$ at time $t$ is not produced instantaneously by the
incoming field at $(x,0)$, but arises from previously excited
positions $(x',0)$ with $x'<x+x''_-$.

It is interesting to note that, according to Eq. (\ref{lorentz}), the
height of the transmitted pulse is inversely proportional to
the distance $z$ from the surface, instead of decaying exponentially
as usually found for evanescent waves. However, Eq. (\ref{lorentz})
describes the propagation into vacuum of a single infinitely sharp
incident wavefront. In a wavetrain made up of a succession of incident
pulses, the regions excited by neighboring pulses overlap each other,
as the width of each transmitted pulse increases in
proportion to $z$, and therefore their corresponding fields
interfere. This interference is at 
the origin of the 
exponential decay of periodic waves, as can be verified by choosing
\begin{equation}\label{periodico}
  f_t(\tau)=A e^{-i\omega \tau}
\end{equation}
and substituting into
Eq. (\ref{phievan}), which yields
\begin{equation}\label{periodic}
\phi(t,x,z)=A e^{i(Qx-\omega t)}\int_{-\infty}^\infty \frac{dt''}{\pi}
e^{-i\omega t''} \frac{\gamma|v_\parallel|z}{z^2+(\gamma v_\parallel
t'')^2}. 
\end{equation}
A simple contour integration closing the integration path with an
infinite semicircle on the lower half complex $t''$ plane yields the
familiar result
\begin{equation}\label{exponential}
\phi(t,x,z)=A e^{i(Qx-\omega t)-\kappa z},
\end{equation}
where $Q=\omega/v_\parallel=n\sin\theta_i\omega/c$ is the parallel
component of the wave vector and
$\kappa=\omega/(\gamma|v_\parallel|)=\sqrt{Q^2-\omega^2/c^2}=1/l$ the
inverse of the decay length $l$. As $\omega$ increases, the distance
along $x$ between successive maxima and minima decreases, yielding
larger interference effects and a shorter decay length.

\section{Screens}\label{secscreens}

\begin{figure}
\includegraphics{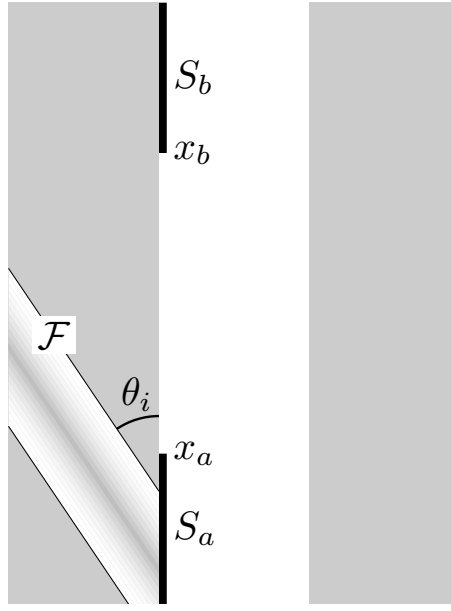}
\caption{\label{screens}Pulse as in Fig. \ref{normal} but impinging on
  the surface of the first dielectric at an angle
  $\theta_i>\theta_c$. The surface is covered by semi-infinite opaque
  screens $S_a$ and $S_b$ with edges at $x_a$ and $x_b$. The leading
  wavefront of the incoming pulse is about to reach $x_a$. $\cal F$
  denotes the foremost wavefront.}
\end{figure}

The results of the previous section suggest an experiment
that could confirm that propagation of evanescent waves in the FTIR
geometry is not superluminal nor acausal. The experiment could be
performed simply by
partially covering the surface of the first interface with a couple of
opaque screens as shown in Fig. \ref{screens}.
If transmission were indeed superluminal, we would expect a non-null
transmitted field $\phi(t,x_a,d)$ across the gap in front of the edge
$x_a$ of the first screen $S_a$ as soon as the leading wavefront $\cal
F$ of the incident pulse reaches $x_a$. Similarly, we would expect
that the field $\phi(t,x_b,d)$ would be modified as soon as $\cal F$
reaches the edge $x_b$ of the second screen $S_b$. 

To calculate the field corresponding to Fig. \ref{screens} we go back
to Eq. (\ref{psievan}). The screens confine the integration region to
the interval $x_a<x'=x+x''<x_b$. This inequality has to be obeyed
together with the previous constriction $x'<x+x''_-$. These conditions
can only be satisfied by those wavefronts which have reached $x_a$ and
left behind the first screen before $t-T_a$, where
$T_i=(1/c)\sqrt{(x-x_i)^2+z^2}$, $i=a,b$, is the minimum time required
to reach $(x,z)$ from $(x_i,0)$ moving at speed $c$. Thus, only those
points on the wavefront labeled by $t''>T''_a$ can contribute to
(\ref{psievan}), where
\begin{equation}\label{T''}
T''_i\equiv
\frac{x-x_i}{v_\parallel}-\frac{1}{c}\sqrt{(x_i-x')^2+z^2},\quad i=a,b.
\end{equation}
For those wavefronts which have left $S_a$ by time $T_a$
but have not reached $S_b$ at time $T_b$, namely, those with
$T''_b<t''<T''_a$, the integral over $x''$ in Eq. (\ref{psievan}) has
to be performed from $x''_a=x+x_a$ up to $x''_-$. Finally, for those
wavefronts which have already been blocked by $S_b$ by time $T_b$,
namely, those with $t''<T''_b$, the upper limit of integration has to
be replaced by $x''_b=x+x_b$. Therefore,
\begin{equation}\label{psiscreens}
\psi(t,x,z)=\frac{\gamma\beta c}{\pi}\int_{T''_b}^{T''_a} dt''\,
f_t(t-x/v_\parallel+t'') \int_{\eta_a}^{-1}
\frac{d\eta}{\sqrt{\eta^2-1}} + \frac{\gamma\beta
  c}{\pi}\int_{-\infty}^{T''_b} dt''\, f_t(t-x/v_\parallel+t'')
\int_{\eta_a}^{\eta_b} \frac{d\eta}{\sqrt{\eta^2-1}}
\end{equation}
where we used the change of variables (\ref{etaagain}) and substituted
$x''\to x+x_i$ in it to define the limits $\eta_i$. Notice that $\psi$
depends on $z$ 
only through the integration limits $\eta_i$, so that substituting
Eq. (\ref{psiscreens}) in (\ref{phivspsi}) we obtain
\begin{equation}\label{psivsxi}
\psi(t,x,z)=\frac{\gamma\beta c}{\pi}\int_{-\infty}^\infty dt''\,
f_t(t-x/v_\parallel+t'') [\xi_a\Theta(T''_a-t'') - \xi_b
  \Theta(T''_b-t'')],
\end{equation}
where
\begin{equation}\label{xi}
\xi_i\equiv
\frac{1}{\sqrt{\eta_i^2-1}}\frac{\partial\eta_i}{\partial
z}\equiv \xi_L \zeta_i,
 \end{equation}
$\xi_L$ is given by Eq. (\ref{detadz})and $\zeta_i\equiv\zeta(x''_i,t'')$ with
\begin{equation}\label{zeta}
\zeta(x'',t'')=-\,\frac{x''+\gamma^2\beta c
t''}{\sqrt{(x'')^2+2\gamma^2 
\beta c x'' t'' - \gamma^2\beta^2 (z^2-c^2 (t'')^2)}}.
\end{equation}
Substituting Eqs. (\ref{etaagain}) and (\ref{xi}) in (\ref{psivsxi})
we finally obtain
\begin{equation}\label{phiscreened}
\phi(t,x,z)=\frac{1}{\pi}\int_{-\infty}^\infty dt''\, 
f_t(t-x/v_\parallel+t'') \frac{\gamma v_\parallel z}{z^2+(\gamma
v_\parallel t'')^2} (1-C(t'')),
\end{equation}
where 
\begin{equation}\label{C}
C(t'')=1+[\zeta_b\Theta(T''_b-t'') - \zeta_a \Theta(T''_a-t'')].
\end{equation}
Notice that the field $\phi$ in the presence of screens
(Eq. (\ref{phiscreened})) is given by an expression similar to that
corresponding to the field in the absence of screens
(Eq. (\ref{phievan})) but with a correction term $C$ due to the
diffraction by the screen.

As in the previous section, we consider again the case of a sharp
incident pulse  Eq. (\ref{sharp}). Substituting in (\ref{phiscreened})
we obtain
\begin{equation}\label{lorentzscreened}
\phi(t,x,z)=\frac{\gamma v_\parallel z}{z^2+\gamma^2(x-v_\parallel t)^2}
\times \left\{
\begin{array}{ll}
0&\mbox{if $ct<x_a/\beta+\sqrt{(x-x_a)^2+z^2}$},\\
\zeta(x_a-x,x/v_\parallel-t)-\zeta(x_b-x,x/v_\parallel-t)&\mbox{if
$ct>x_b/\beta+\sqrt{(x-x_b)^2+z^2}$},\\ 
\zeta(x_a-x,x/v_\parallel-t)&\mbox{otherwise}.
\end{array}
\right.
\end{equation}
We remark that the field is zero until the time $x_a/v_\parallel$ when
the incident pulse shows up from behind the screen $S_a$, and this
information has had enough time $(1/c)\sqrt{(x_a-x)^2+z^2}$ to
propagate from the screen's edge $(x_a,0)$ to the observation point
$(x,z)$. Similarly, the information that the pulse has hidden behind
screen $S_b$ does not reach the observation point until the time
$x_b/v_\parallel+(1/c)\sqrt{(x-x_b)^2+z^2}$. The field has
singularities due to the passage of the incident pulse through the
screen edges, that propagate at speed $c$ from the events
$(x_a/\beta,x_a,0)$ and $(x_b/\beta,x_b,0)$. Notice that
$\zeta(x_i-x,0)\to 1$ as $x\to\infty$. Thus, if we follow the incident
pulse, i.e., we take $x\approx v_\parallel t$, then $\phi\to 0$
asymptotically after the pulse hides behind $S_b$. Furthermore, if the
screens are very far apart we recover the field (\ref{lorentz})
between the screens.

\begin{figure}
\includegraphics{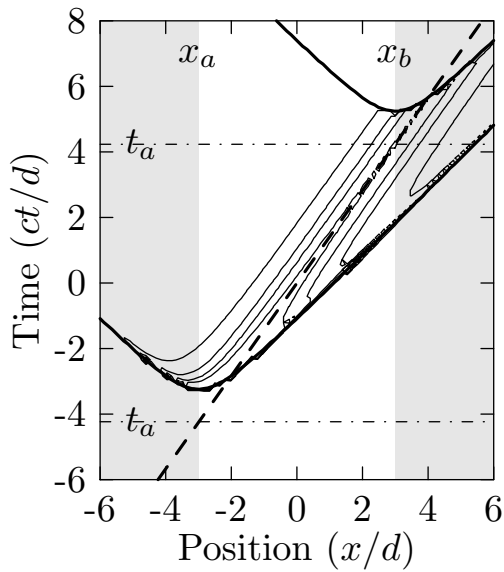}
\caption{\label{pantalla}Field isolines of $\phi(t,x,d)$ produced by
  an unit delta function input pulse that propagates along the $z=0$
  surface with speed $v_\parallel=c/\sqrt{2}$ and is blocked at
  $x<x_a=-3d$ and $x>x_b=3d$ by opaque screens (shaded regions). The
  world line $x=v_\parallel t$ of the incident pulse is indicated by a
  dashed line. The times $t_a$ and $t_b$ when the incident pulse cross
  the edges of each screen are indicated by the horizontal dot-dashed
  lines. The singularities of the transmitted field are indicated by
  the thick solid hyperbolas.}
\end{figure}

The features above are illustrated in Fig. \ref{pantalla}
which show the transmitted field at the plane $z=d$. Notice the delay
$d/c$ after the incident pulse crosses $x_a$ at $t_a=x_a/v_\parallel$
before a non-null field first appears across the gap at $(x_a,d)$, and
a similar delay after the incident pulse crosses $x_b$ at
$t_b=x_b/v_\parallel$ before the field starts to be extinguished at
$(x_b,d)$. Furthermore, notice that for some time the field penetrates
a small distance $\approx d$ beyond $x_b$ as if there were no
screen. The field is singular at the hyperbolas with vertices at
$x=x_a$, $t=t_a+d/c$ and at $x=x_b$, $t=t_b+d/c$ given by the
intersection of the $(ct,x,d)$ hyperplane and the future light cone of
the events $(c t_a,x_a,0)$, $(c t_b,x_b,0)$. Thus, we have shown
observable consequences of the fact that evanescent waves in FTIR do
not propagate superluminally nor acausally in the direction normal to
the dielectric-vacuum interfaces, but with retardation and
obliquely. A graphical approach to the results of this section and an
animation illustrating them may be found in
Refs. \onlinecite{mochan01} and Ref. \onlinecite{oe} respectively.

\section{A smooth transverse profile}\label{secsmooth}

Perfectly opaque screens such as those considered in the previous
section introduce sharp discontinuities in the pulse at $z=0$. The
truncated pulse no longer has a well defined propagation direction
$\theta_i$ but may still be represented by a superposition of pulses
with varying propagation directions. A sharp truncation leads to the
presence of subcritical incident angles $\theta_i<\theta_c$, and
therefore to the presence of both, evanescent and non-evanescent
transmitted fields. It has been argued \cite{carey01} that the
retardation effects discussed in the previous section may be due only
to the non-evanescent contributions,  known to be
subluminal.  The comparatively slow subluminal contributions would be
unable to affect the arrival of the superluminal signals if the later
were actually present. However, any small non-evanescent wave would
dominate the transmitted signal after a wide enough gap. Thus, it is
interesting to study the propagation of pulses with a finite
transverse extension but with a smooth lateral cutoff and built up
completely from hypercritical $\theta_i>\theta_c$ evanescent
contributions.

To explore the propagation of the smoothly truncated pulses discussed
above, we consider an incoming field given by a Fourier integral
\begin{equation}\label{fourier}
\phi(t,x,0^+)=\int\frac{d\omega}{2\pi}\int\frac{d Q}{2\pi}\, f_{\omega Q}
e^{i(Q x - \omega t)},
\end{equation}
where $f_{\omega Q}$ is the amplitude for each parallel component of
the wave vector $Q$ and frequency $\omega$. We can change integration
variable from $Q$ to the parallel velocity $v\equiv \omega/Q$,
\begin{equation}\label{vfourier}
\phi(t,x,0^+)=\int\frac{d\omega}{2\pi}\int\frac{d
v}{2\pi}\, f_{\omega v} 
e^{-i\omega (t-x/v)}.
\end{equation}
where we introduced the velocity dependent amplitude $f_{\omega v}
\equiv (\omega/v^2) f_{\omega,\omega/v}$.  The incident field
(\ref{vfourier}) will give rise to evanescent waves exclusively as
long as all non-null components $f_{\omega v}$ have $v<c$.  At this
point we could integrate first Eq. (\ref{vfourier}) with respect to
$\omega$, obtaining thus a superposition of plane pulses, each of
which may be propagated across the air gap according to
Eq. (\ref{phievan}). Alternatively, we may propagate each
monochromatic component using Eq. (\ref{exponential}) and afterward
perform the integrations in Eq. (\ref{vfourier}). We follow the later
approach and write
\begin{equation}\label{expprop}
\phi(t,x,z)=\int\frac{d\omega}{2\pi}\int\frac{d v}{2\pi}\, f_{\omega
v} e^{-\omega[z/(\gamma v)+i(t-x/v)]}, 
\end{equation}
where $\gamma=1/\sqrt{1-v^2/c^2}$ as in Sec. \ref{secevan}. For
simplicity we assume that we may factor $f_{\omega v}=f_\omega f_v$
into frequency and velocity dependent amplitudes, $f_\omega$ and
$f_v$ respectively; $f_\omega$ controls the time duration of the
pulse, or equivalently, its longitudinal extent, while $f_v$ controls
its transverse extent. We further assume a narrow Gaussian velocity
distribution of width $\Delta v$ around a nominal velocity $v_0<c$, 
\begin{equation}\label{gaussian}
f_v= \frac{\sqrt{2\pi}}{\Delta v} e^{-u^2/2\Delta v},
\end{equation}
where $u\equiv v-v_0$. We assume $\Delta v$ is small enough that the
non-evanescent contributions to the field may be neglected and the
exponent in Eq. (\ref{expprop}) may be linearized in $u$. Thus, the
transmitted field becomes 
\begin{equation}\label{transm}
\phi(t,x,z)\approx \int\frac{d\omega}{2\pi}\, f_\omega
e^{-\omega[z/(\gamma_0 v_0)+i(t-x/v_0)]} \int\frac{d
u}{\sqrt{2\pi}\Delta v}\,  e^{-u^2/2\Delta v^2 +\omega u[z/\gamma_0
-i x]/v_0^2}, 
\end{equation}
where $\gamma_0=1/\sqrt{1-v_0^2/c^2}$. The integration over $u$
is immediate and yields
\begin{equation}\label{transm1}
\phi(t,x,z)= \int\frac{d\omega}{2\pi}\, f_\omega
e^{-\omega[z/(\gamma_0 v_0)+i(t-x/v_0)] - [\Delta v \omega(x+i
z/\gamma_0)]^2/2 v_0^4}.
\end{equation}

\begin{figure}
\includegraphics{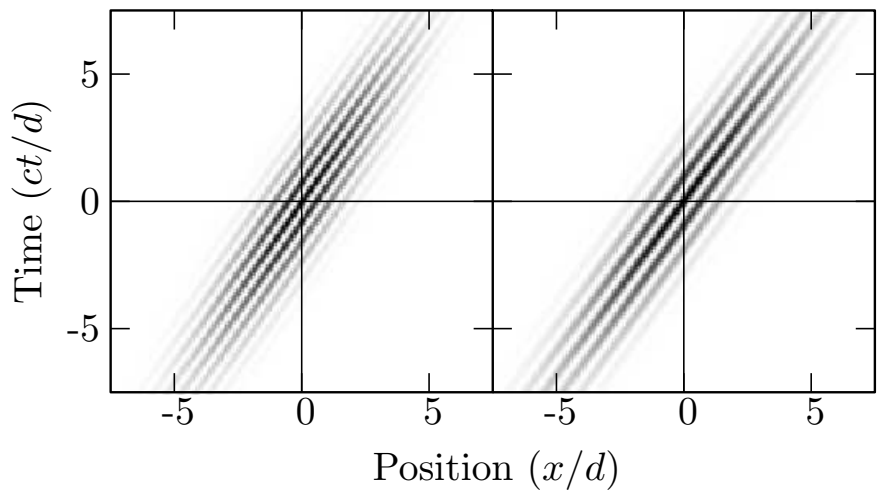}
\caption{\label{figgauss}Intensity of a Gaussian pulse localized in
both space and time ($x$ and $t$) incident on the front face of an air
gap of width $d$ (left) and after after crossing it (right). The
nominal velocity along $x$ is $v_0=0.7c$ with width $\Delta
v=0.15c$. The nominal frequency is $\omega_0=16 c/d$ with width $\Delta
\omega =2 c/d$}
\end{figure}

In Fig. \ref{figgauss} we illustrate the results of applying
Eq. (\ref{transm1}) to a pulse with a Gaussian frequency
distribution,
\begin{equation}
f_\omega = A \frac{\sqrt{2\pi}}{\Delta\omega}
e^{-(\omega-\omega_0)^2/2\Delta\omega^2},
\end{equation}
of area $A$ and width $\Delta\omega$ centered at $\omega_0$
($\omega_0=16c/d$, $\Delta\omega=2c/d$, $v_0=0.7c$, $\Delta v=0.15c$).
The pulse is seen to form on the $z=0$ surface at $x\approx -5d$ and
at time $t\approx7d/c$, it propagates along the $x\approx v_0 t$ line
for a while, peaks at $x=0$ at time $t=0$ and disappears at $x\approx
5d$, $t\approx7d/c$. Its maximum duration $\tau$ and size $L$ are
$c\tau\approx L\approx d$ for a fixed position and fixed observation
time respectively, and it contains altogether about six nodes.

Surprisingly, after crossing the gap, the pulse looks essentially the
same! It appears at the back  $z=d$ face of the air gap at roughly the
same time and the same position as at the front $z=0$ face. It also
peaks at the origin at $t=0$ and disappears from the back surface in
concordance to the incident pulse on the front face. Thus, it truly
appears to propagate instantaneously. The main difference between the
incident and transmitted pulse is that the intensity of the later is
suppressed by 12 orders of magnitude. Another interesting difference is
that the number of visible nodes in the transmitted pulse has
decreased to about 4. This is a consequence of the fact that in FTIR,
the Fourier components with higher frequencies are more damped than
those with lower frequencies. Finally, a more subtle difference is
that the speed of propagation along the back face is slightly but
noticeably larger than that on the front face. This is due to the fact
that plane waves incident at angles closer to $\theta_c$ have a larger
penetration length than waves incident at larger angles. Thus, the
angle of propagation of the transmitted pulse is smaller than that of
the 
incident wave \cite{balcou97}.

\begin{figure}
\includegraphics{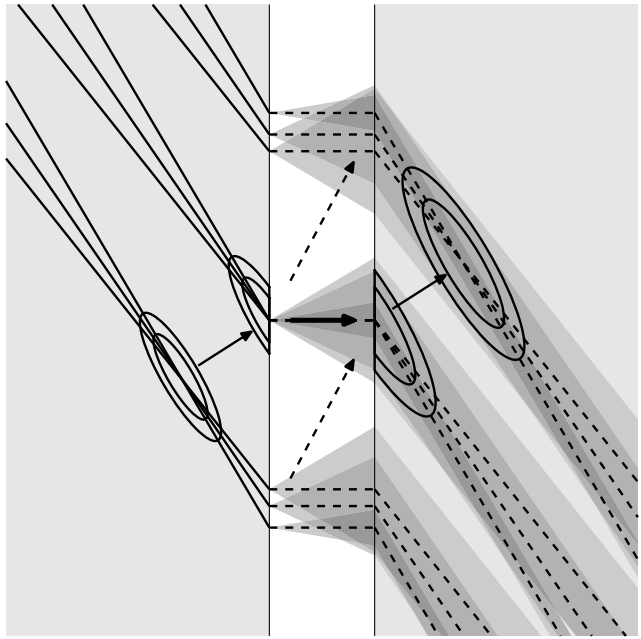}
\caption{\label{misaligned}Evanescent transmission of a smoothly
truncated pulse made up initially of a
superposition of narrow plane wavefronts (heavy solid lines) with a
distribution of angles above 
$\theta_c$. As it crosses the gap, each wavefront widens (shaded bands
around dashed lines, as in Fig. \ref{figlorentz}). The ellipses
represent 
schematically the 
contour levels of 
the pulse and are centered at the region where the different
contributions add coherently in phase, i.e., the regions where the
centers of each component coincide. Snapshots are taken at three
different times. The solid arrows illustrate the nominal propagation
of the peak. The heavy arrow indicates the apparent instantaneous
tunneling across the air gap. 
The dashed arrows illustrate the actual (sub)luminal propagation.} 
\end{figure}

In Sec. \ref{secevan} we have argued that the transmission of
evanescent plane pulses across an air gap under FTIR conditions is
fully consistent with a retarded and causal propagation along oblique
directions. We have strengthened our argument by showing that there is
a
delay before a perturbation, such as blocking part of the incident
wavefront, can produce an effect on the pulse transmitted across the
gap. Furthermore, by truncating an incident plane pulse 
producing an abrupt transverse profile, we showed that the transmitted
pulse is shifted along the surface in the direction of propagation. 
However, in Sec. \ref{secsmooth} we showed through an example that if
the pulse has a smooth transverse profile, such that all its
Fourier components are evanescent, it is transmitted as if it were
indeed 
superluminal. To understand this result, in Fig. \ref{misaligned} we
show schematically a pulse  smoothly truncated along its transverse
direction, built from narrow plane components propagating along well
defined directions $\theta>\theta_c$. 
The incident field peaks at the regions
where the directional components add in phase, indicated in the figure
by elliptical regions around the crossing point of the different
incident wavefronts. As each of the components crosses the gap,
it is widened according to Eq. (\ref{lorentz}). The peak of the
transmitted pulse appears in the regions of largest overlap between
the different widened transmitted components, i.e., at the regions
where their centers coincide. The peak of the transmitted pulse
at the back face of the air gap is seen to appear at the same time as
the peak of 
the incident pulse reaches the front surface. Thus, it would seem as
if the peak 
tunneled instantaneously in the direction normal to the gap. However,
in the previous sections we have shown that each of the
transmitted components of the transmitted wave originates causally
from regions in the lateral wings of the incident pulse. Thus, the
peak of the transmitted field does not actually originate from the
peak of the incident field; it is formed by
contributions from the lateral wings of the incident field, which
reach the front face of the air gap first. The lateral wings of each
component have enough time to cross the gap traveling at speed
$c$ and combine to form the relatively small transmitted peak right at
the time when 
the larger incident peak reaches the front surface. Similarly, the
different 
components of the field produced by the peak of the incident pulse get
out of step as they cross the air gap and, therefore, do not
contribute to the peak of the transmitted pulse, but rather, to its
lateral wing. This is illustrated by the dashed arrows in
Fig. \ref{misaligned}. Thus, it seems that the physical
propagation (retarded and  causal) can not be
distinguished from the nonphysical propagation (superluminal and
non-causal) as long as we only consider smooth incident pulses which
contain only propagation directions above the critical angle
\cite{carey00,carey01}. 

\section{Evanescent propagator}\label{secevprop}

In Sec. \ref{secsmooth} we found that a particular pulse made up of only
evanescent components seemed to propagate instantaneously across the
air gap, in contrast to the abruptly truncated pulses considered in
Sec. \ref{secscreens} for which retardation effects have observable
consequences. This was explained graphically in  Fig. \ref{misaligned}
for incident pulses built up from narrow
directional components, each of which is widened as it is transmitted
across the gap. To show that this behavior is generic, we start from
the Fourier decomposition of an arbitrary field
\begin{equation}\label{Fourier}
\phi(t,x,z)=\int\frac{d Q}{2\pi} \int\frac{d\omega}{2\pi}
e^{i(Q x - \omega t)} \phi_{\omega,Q}(z), 
\end{equation}
where
\begin{equation}\label{Fourierinv}
\phi_{\omega,Q}(z)=\int d t\int d x\, e^{-i(Q x - \omega t)}
\phi(t,x,z). 
\end{equation}
The condition that $\phi(t,x,z)$ is made up exclusively of evanescent
waves is equivalent to stating that the integration region in
Eq. (\ref{fourier}) is given by $|\omega|<|Q|c$, i.e.,
$\phi_{\omega,Q}=0$ 
if $|\omega|>|Q|c$. Using Eq. (\ref{exponential}) we propagate  each
Fourier component from $z=0^+$ to $z>0$ as 
\begin{equation}\label{phiwQz}
\phi_{\omega, Q}(z)=e^{-\kappa z} \phi_{\omega, Q}(0^+),
\end{equation}
where $\kappa=\sqrt{Q^2-\omega^2/c^2}$. Thus, we can combine
Eqs. (\ref{Fourier}), (\ref{Fourierinv}) and (\ref{phiwQz}) to obtain
\begin{equation}\label{propagate1}
\phi(t,x,z)=\int dx' \int dt'\,P'(t,x,z;t',x',0^+) \phi(t',x',0^+),
\end{equation}
where
\begin{equation}\label{propagator1}
P'(t,x,z;t',x',0^+)\equiv P'(t-t',x-x',z)=\int\frac{dQ}{2\pi}
\int\frac{d\omega}{2\pi} e^{i[Q(x-x')-\omega(t-t')]-\kappa
  z} 
\end{equation}
and the integration region is given by $|\omega|<|Q|c$.
Comparing Eq. (\ref{propagate1}) with (\ref{propagate}) we find that
$P'$ is a propagator that can be used in the same way as the
propagator $P$ defined in Eq. (\ref{P})  
to find the value of
the field at 
$z>0$ given its values at $z=0^+$, provided the field is built up of evanescent
components only, as in the examples of the two previous sections. We
remark that our original causal, retarded and subluminal propagator
$P$ was able to propagate any arbitrary outgoing field. However,
by adding constrains to the field, we gain freedom in our choice of
propagator, as we can chose arbitrarily its effect on fields that do
not obey the constrain. Thus, if we can find any function $P''$ such that
\begin{equation}\label{constrain}
\int dx'\int dt'\,[P''(t,x,z;t',x',0^+)-P(t,x,z;t',x', 0^+)] f(t'-x'/v) =0,
\end{equation}
for an arbitrary flat pulse $f$ moving along $x$ with any velocity
$-c<v<c$, then we could employ $P''$ instead of $P$ to propagate an
arbitrary evanescent pulse. $P'$ above is just one of the
many possible choices of a propagator for evanescent pulses.

To proceed, we make a change of variable $\omega\to Qv$ to write
Eq. (\ref{propagator1}) as
\begin{equation}\label{prop2}
P'(\tau,\xi,z)=\int\frac{dQ}{2\pi}\int_{-c}^c\frac{dv}{2\pi}\, |Q| e^{iQ(\xi-v\tau)}
e^{-Qz/\gamma},
\end{equation}
where $\gamma=1/\sqrt{1-v^2/c^2}$, $\xi=x-x'$, $\tau=t-t'$ and we
perform the integration over 
$Q$,
\begin{equation}\label{prop3}
P'(\tau,\xi,z)=\frac{1}{2\pi^2} \int_{-c}^c
dv\,\frac{(z/\gamma)^2(\xi-v\tau)^2} 
{[(z/\gamma)^2+(\xi-v\tau)^2]^2}.
\end{equation}
Notice that $P'(t,x,z;t',x',0^+)$ is symmetric under the interchange
$\xi\leftrightarrow -\xi$ and also under the interchange 
$\tau\leftrightarrow -\tau$, i.e., $P'(t,x,z;t',x') = P'(t,x',z;t',x) =
P'(t',x,z;t,x') = P'(t',x',z;t,x)$. Thus, the evanescent propagator
$P'$ {\em is superluminal and acausal}.

To finish the calculation of $P'$ we make another change of
integration variable $v=c \sin\alpha$ to write
\begin{equation}\label{prop4}
P'(\tau,\xi)=\frac{c}{2\pi^2}\int_{-\pi/2}^{\pi/2} d\alpha \cos\alpha
\frac{(z\cos\alpha)^2-(\xi-c \tau \sin\alpha)^2}
{[(z\cos\alpha)^2+(\xi-c \tau \sin\alpha)^2]^2},
\end{equation}
and we perform the integration
\begin{equation}\label{evprop}
P'(\tau,\xi)=\frac{c}{4\pi^2}\frac{1}{s^3} \left[ z \log
\left( \frac{(c^2 \tau^2-z^2-z s)^2-c^2 \tau^2
  \xi^2} {(c^2 
\tau^2-z^2+z s)^2-c^2 \tau^2 \xi^2}\right)
-4 s\right],
\end{equation}
where
$\rho=\sqrt{\xi^2+z^2}$ 
is the spatial distance from the source to the
observation point and $s^2 = \rho^2-c^2 \tau^2$ is the squared
space-time interval.
This expression may be simplified to
\begin{equation}\label{evprop1}
P'(\tau,\xi)= \frac{c}{2\pi^2}\times \left\{ 
\begin{array}{ll}
 z \log(|z+s|/|z-s|)/s^3 -2/s^2 &\mbox{if $s^2 > 0$},\\
-2z\arctan(|s|/z)/|s|^3-2/|s|^2
&\mbox{if $s^2 < 0$},\\
\end{array}
\right..
\end{equation}

\begin{figure}
\includegraphics{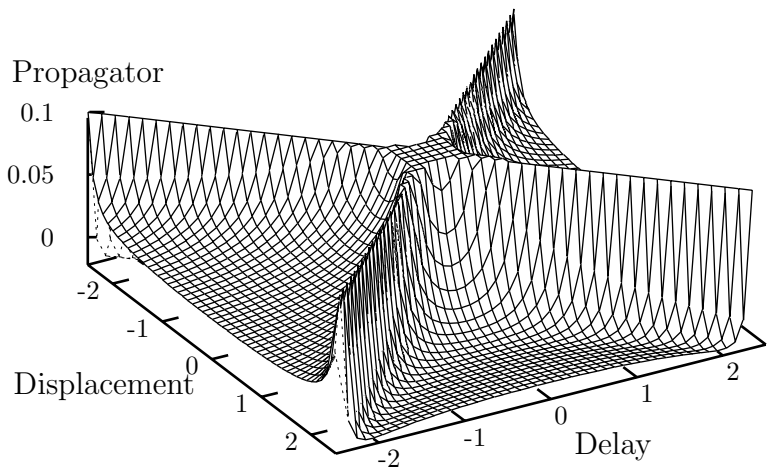}
\caption{\label{figpropagador} Propagator $P'(t,x,d;t',x',0^+)$ normalized
to $c/d^2$ as a
function of the displacement $x-x'$ and the delay $t-t'$. Distance is
measured in units of $d$ and time in units of $d/c$. For aid in
visualization, the height of the propagator was truncated at $P'=0.1
c/d^2$. }
\end{figure}

The evanescent propagator is displayed in Fig. \ref{figpropagador}. 
The figure shows explicitly the temporal and spatial symmetry, and
thus the superluminality and non-causality of $P'$. Notice that $P'$
has singularities at the projected light-lines $x-x'=\pm c(t-t')$ which
converge at the origin $x=x'$, $t=t'$. Thus, propagation is largest
for instantaneous propagation in the direction normal to the air
gap. 

We have found that the propagation of evanescent pulses can be
described with either the exact propagator $P$ of the problem, which is
causal, retarded and subluminal, or with an {\em evanescent}
propagator $P'$ which is superluminal and non-causal. Both yield
exactly the same transmitted pulse when the incident pulse contains
only evanescent components. Thus, it seems to be impossible to
distinguish superluminal from subluminal propagation in experiments
performed with purely evanescent pulses \cite{carey01}.

An explanation for the curious conclusion found in this section can be
obtained by going back to Eq. (\ref{Fourier}) which we rewrite as
\begin{equation}
\phi(t,x,z)=\int\frac{d Q}{2 \pi} \phi_Q(t,z).
\end{equation}
Notice that for each
finite wave vector $Q$, the time dependent Fourier component
$\phi_Q(t,z)$ has a strictly finite spectrum
$-|Q|c<\omega<|Q|c$. Thus, $\phi_Q(t,z)$ is an analytical function of
$t$ with no singularities and can be analytically
continued to an arbitrary time $t_2$ from its values in an arbitrarily small
neighborhood of any other arbitrary time $t_1$. Therefore,
$\phi_Q(t,z)$ is perfectly 
{\em predictable} in principle. Using antropomorphic language, we may
say that at time $t_1$ the system {\em knows} from
the present values of $\phi(t\approx t_1,x,0^+)$ what its future values
$\phi(t\approx t_2 \ge t_1+z/c, x, 0^+)$ will be, and thus it 
can {\em use} that knowledge to build subluminally a transmitted pulse
$\phi(t\approx t_2,x,z)$ that will mimic $\phi(t\approx t_2, x, 0^+)$,
giving the impression that superluminal transmission has taken
place at time $t_2$. The validity of the argument above in the
presence of thermal or quantum noise has to be investigated.

\section{Conclusions}\label{secconclusion}

To study the propagation of a light pulse through a vacuum gap between
two parallel dielectrics in a FTIR configuration we constructed a propagator
derived directly from the wave equation resulting from Maxwell's
equations. This  
propagator is retarded and complies with the relativistic causality
principle inherent to classical electromagnetism. Therefore, it can
only account for {\em causal, (sub)luminal}  
propagation of light pulses. However, when this propagator is used to
study the propagation of wave packets through the gap, we find
apparent superluminal behavior, that is, a wave packet might appear on
the far side of the gap at the same time that the incident packet
reaches the front one.  Therefore this illusion of
superluminality, present within classical electromagnetic
theory even in vacuum, is fully consistent with relativistic
causality. We showed explicitly that propagation  in FTIR actually takes place
subluminally between the lateral wings of the incident pulse and the
central peak  of the transmitted pulse, and we proposed simple
experiments that could verify this statement. Thus, although FTIR has
many similitudes to 1D tunneling, its correct physical interpretation
requires a 2D or 3D analysis. On the other hand, we
constructed an explicitly superluminal and acausal propagator that
yields identical results as the retarded causal one when applied to
smooth pulses made up of evanescent contributions only. Thus, there is
a class of pulses for which superluminal and subluminal propagation
would be indistinguishable. 

\begin{acknowledgments}
We acknowledge partial support from 
UBACYT and CONICET (VLB) and from DGAPA-UNAM under
project IN111306 (WLM). VLB is a member of CONICET.  
\end{acknowledgments}


\begin{thebibliography}{7}
\expandafter\ifx\csname natexlab\endcsname\relax\def\natexlab#1{#1}\fi
\expandafter\ifx\csname bibnamefont\endcsname\relax
  \def\bibnamefont#1{#1}\fi
\expandafter\ifx\csname bibfnamefont\endcsname\relax
  \def\bibfnamefont#1{#1}\fi
\expandafter\ifx\csname citenamefont\endcsname\relax
  \def\citenamefont#1{#1}\fi
\expandafter\ifx\csname url\endcsname\relax
  \def\url#1{\texttt{#1}}\fi
\expandafter\ifx\csname urlprefix\endcsname\relax\def\urlprefix{URL }\fi
\providecommand{\bibinfo}[2]{#2}
\providecommand{\eprint}[2][]{\url{#2}}



\bibitem[{\citenamefont{Wang et~al.}(2007)\citenamefont{Wang and Xiong}}]{wang07}
\bibinfo{author}{\bibfnamefont{Zhi-Yong}~\bibnamefont{Wang}} \bibnamefont{and}
  \bibinfo{author}{\bibfnamefont{Cai-Dong}~\bibnamefont{Xiong}},
  \bibinfo{journal}{Phys. Rev. A} \textbf{\bibinfo{volume}{75}},
  \bibinfo{pages}{042105} (\bibinfo{year}{2007}).

\bibitem[{\citenamefont{Winful}(2007)\citenamefont{Winful}}]{winful07}
 \bibinfo{author}{\bibfnamefont{H. G.}~\bibnamefont{Winful}},
  \bibinfo{journal}{Phys. Rev. A} \textbf{\bibinfo{volume}{76}},
  \bibinfo{pages}{057803} (\bibinfo{year}{2007}).

\bibitem[{\citenamefont{Pereyra et~al.}(2007)\citenamefont{Pereyra and Simanjuntak}}]{pereyra}
\bibinfo{author}{\bibfnamefont{P.}~\bibnamefont{Pereyra}} \bibnamefont{and}
  \bibinfo{author}{\bibfnamefont{H.P.}~\bibnamefont{Simanjuntak}},
  \bibinfo{journal}{Phys. Rev. E} \textbf{\bibinfo{volume}{75}},
  \bibinfo{pages}{056604} (\bibinfo{year}{2007}).

\bibitem[{\citenamefont{Ranfagni et~al.}(2007)\citenamefont{Ranfagni et al}}]{ranfagni07}
\bibinfo{author}{\bibfnamefont{A.}~\bibnamefont{Ranfagni}},
  \bibinfo{author}{\bibfnamefont{G.}~\bibnamefont{Viliani}},
  \bibinfo{author}{\bibfnamefont{C.}~\bibnamefont{Ranfagni}},
  \bibinfo{author}{\bibfnamefont{R.}~\bibnamefont{Mignani}},
  \bibinfo{author}{\bibfnamefont{R.}~\bibnamefont{Ruggeri}} \bibnamefont{and} 
  \bibinfo{author}{\bibfnamefont{A.M.}~\bibnamefont{Ricci}},
  \bibinfo{journal}{Phys. Lett. A} \textbf{\bibinfo{volume}{370}},
  \bibinfo{pages}{370} (\bibinfo{year}{2007}).


\bibitem[{\citenamefont{Winful}(2005)\citenamefont{Winful}}]{winful05}
\bibinfo{author}{\bibfnamefont{H. G.}~\bibnamefont{Winful}},
  \bibinfo{journal}{Phys. Rev. E} \textbf{\bibinfo{volume}{72}},
  \bibinfo{pages}{046608} (\bibinfo{year}{2005}).

\bibitem[{\citenamefont{Winful}(2003)\citenamefont{Winful}}]{winful03}
\bibinfo{author}{\bibfnamefont{H. G.}~\bibnamefont{Winful}},
  \bibinfo{journal}{Phys. Rev. Lett} \textbf{\bibinfo{volume}{90(2)}},
  \bibinfo{pages}{23901} (\bibinfo{year}{2003}).


\bibitem[{\citenamefont{Buttiker et~al.}(2003)\citenamefont{Buttiker and
  Washburn}}]{buttiker03}
\bibinfo{author}{\bibfnamefont{M.}~\bibnamefont{Buttiker}} \bibnamefont{and}
  \bibinfo{author}{\bibfnamefont{S.}~\bibnamefont{Washburn}},
  \bibinfo{journal}{Nature} \textbf{\bibinfo{volume}{422}},
  \bibinfo{pages}{271} (\bibinfo{year}{2003}).


\bibitem[{\citenamefont{Windul}(2003-1)\citenamefont{Winful}}]{winful03-1}
\bibinfo{author}{\bibfnamefont{Herbert G.}~\bibnamefont{Winful}},
  \bibinfo{journal}{Nature} \textbf{\bibinfo{volume}{424}},
  \bibinfo{pages}{628} (\bibinfo{year}{2003}).

\bibitem[{\citenamefont{Buttiker et~al.}(2003)\citenamefont{Buttiker and  Washburn}}]{buttiker03-1}
\bibinfo{author}{\bibfnamefont{M.}~\bibnamefont{Buttiker}} \bibnamefont{and}
  \bibinfo{author}{\bibfnamefont{S.}~\bibnamefont{Washburn}},
  \bibinfo{journal}{Nature} \textbf{\bibinfo{volume}{424}},
  \bibinfo{pages}{638} (\bibinfo{year}{2003}).

\bibitem[{\citenamefont{Winful}(2003-2)\citenamefont{Winful}}]{winful03-2}
\bibinfo{author}{\bibfnamefont{H. G.}~\bibnamefont{Winful}},
  \bibinfo{journal}{Phys. Rev. E} \textbf{\bibinfo{volume}{68}},
  \bibinfo{pages}{016615} (\bibinfo{year}{2003}).

\bibitem[{\citenamefont{Peres et~al.}(2004)\citenamefont{Peres and Terno}}]{qi}
\bibinfo{author}{\bibfnamefont{A.}~\bibnamefont{Peres}} \bibnamefont{and}
  \bibinfo{author}{\bibfnamefont{D.}~\bibnamefont{Terno}},
  \bibinfo{journal}{Rev. Mod. Phys.} \textbf{\bibinfo{volume}{76}},
  \bibinfo{pages}{93} (\bibinfo{year}{2004}).

\bibitem[{\citenamefont{De Angelis et~al}(2007)\citenamefont{De Angelis et al}}]{de_angelis}
\bibinfo{author}{\bibfnamefont{T.}~\bibnamefont{De Angelis}},
\bibinfo{author}{\bibfnamefont{E.}~\bibnamefont{Nagali}},
\bibinfo{author}{\bibfnamefont{F.}~\bibnamefont{Sciarrino}} \bibnamefont{and}
  \bibinfo{author}{\bibfnamefont{F.}~\bibnamefont{De Martini}},
  \bibinfo{journal}{Phys. Rev. Lett.} \textbf{\bibinfo{volume}{99}},
  \bibinfo{pages}{193601} (\bibinfo{year}{2007}).


\bibitem[{\citenamefont{Winful}(2006)\citenamefont{Winful}}]{winful06}
 \bibinfo{author}{\bibfnamefont{H. G.}~\bibnamefont{Winful}},
  \bibinfo{journal}{Physics Reports} \textbf{\bibinfo{volume}{436}},
  \bibinfo{pages}{1} (\bibinfo{year}{2006}).

\bibitem[{\citenamefont{Boyd et~al.}(2002)\citenamefont{Boyd and Gauthier}}]{boyd02}
\bibinfo{author}{\bibnamefont{R. W. Boyd and D.J. Gauthier}}, \emph{\bibinfo{title}{Progress in Optics, Vol. 43, E. Wolf (ed.)}} 
(\bibinfo{publisher}{Elsevier},
  \bibinfo{address}{Amsterdam}, \bibinfo{year}{2002}).



\bibitem[{\citenamefont{Garrison et~al}(1998)\citenamefont{Garrison,Mitchell, Chiao and
 Bolda}}]{garrison1998}
\bibinfo{author}{\bibfnamefont{J. C.}~\bibnamefont{Garrison}},
  \bibinfo{author}{\bibfnamefont{M. W.}~\bibnamefont{Mitchell}},
  \bibinfo{author}{\bibfnamefont{R. Y.}~\bibnamefont{Chiao}} \bibnamefont{and}
  \bibinfo{author}{\bibfnamefont{E. L.}~\bibnamefont{Bolda}},
  \bibinfo{journal}{Phys. Lett. A} \textbf{\bibinfo{volume}{245}},
  \bibinfo{pages}{19} (\bibinfo{year}{1998}).


\bibitem[{\citenamefont{Azbel'}(1994)\citenamefont{Azbel}}]{azbel}
\bibinfo{author}{\bibfnamefont{Mark Ya.}~\bibnamefont{Azbel'}},
  \bibinfo{journal}{Solid State Commun.} \textbf{\bibinfo{volume}{91(6)}},
  \bibinfo{pages}{439} (\bibinfo{year}{1994}).


\bibitem[{\citenamefont{Jackson}(Jackson)}]{jackson}
\bibinfo{author}{\bibnamefont{J. D. Jackson}}, \emph{\bibinfo{title}{Classical Electrodynamics}} 
(\bibinfo{publisher}{Wiley}, \bibinfo{address}{New York},
  \bibinfo{year}{1975}), 
  \bibinfo{edition}{2nd.} ed.

\bibitem[{\citenamefont{Sommerfeld}(1914)\citenamefont{Sommerfeld}}]{sommerfeld}
\bibinfo{author}{\bibfnamefont{A.} \bibnamefont{Sommerfeld}},
    \bibinfo{journal}{Ann. Physik} \textbf{\bibinfo{volume}{44}},
  \bibinfo{pages}{177} (\bibinfo{year}{1914}).
English translation available in Chap. II of \cite{brillouin-wp}.

\bibitem[{\citenamefont{Brillouin1960}(Brillouin)}]{brillouin-wp}
\bibinfo{author}{\bibnamefont{L. Brillouin}}, \emph{\bibinfo{title}{Wave Propagtion and Group Velocity}} 
(\bibinfo{publisher}{Academic Press},
  \bibinfo{address}{New York}, \bibinfo{year}{1960}).

\bibitem[{\citenamefont{Diener}(1996)\citenamefont{Diener}}]{diener96}
  \bibinfo{author}{\bibfnamefont{G.}~\bibnamefont{Diener}},
  \bibinfo{journal}{Phys. Lett. A} \textbf{\bibinfo{volume}{223}},
  \bibinfo{pages}{327} (\bibinfo{year}{1996}).

\bibitem[{\citenamefont{Diener}(1997)\citenamefont{Diener}}]{diener97}
  \bibinfo{author}{\bibfnamefont{G.}~\bibnamefont{Diener}},
  \bibinfo{journal}{Phys. Lett. A} \textbf{\bibinfo{volume}{235}},
  \bibinfo{pages}{118} (\bibinfo{year}{1997}).


\bibitem[{\citenamefont{Kuzmich et~al.}(2000)\citenamefont{Kuzmich, Dogariu, Wang, Milonni and Chao}}]{kuzmich}
\bibinfo{author}{\bibfnamefont{A.}~\bibnamefont{Kuzmich}},
\bibinfo{author}{\bibfnamefont{A.}~\bibnamefont{Dogariu}},
\bibinfo{author}{\bibfnamefont{L.J.}~\bibnamefont{Wang}},
  \bibinfo{author}{\bibfnamefont{P.W.}~\bibnamefont{Milonni}} \bibnamefont{and}
  \bibinfo{author}{\bibfnamefont{R.Y.}~\bibnamefont{Chiao}},
  \bibinfo{journal}{Phys. Rev. Lett.} \textbf{\bibinfo{volume}{86}},
  \bibinfo{pages}{3925} (\bibinfo{year}{2001}).

\bibitem[{\citenamefont{Stenner et~al.}(2003)\citenamefont{Stenner, Gauthier and Neifeld}}]{stenner}
\bibinfo{author}{\bibfnamefont{M.D.}~\bibnamefont{Stenner}},
  \bibinfo{author}{\bibfnamefont{D.J.}~\bibnamefont{Gauthier}} \bibnamefont{and}
  \bibinfo{author}{\bibfnamefont{M.A.}~\bibnamefont{Neifeld}},
  \bibinfo{journal}{Nature} \textbf{\bibinfo{volume}{425}},
  \bibinfo{pages}{695} (\bibinfo{year}{2003}).

\bibitem[{\citenamefont{Stenner et~al.}(2005)\citenamefont{Stenner, Gauthier and Neifeld}}]{stenner05}
\bibinfo{author}{\bibfnamefont{M.D.}~\bibnamefont{Stenner}},
  \bibinfo{author}{\bibfnamefont{D.J.}~\bibnamefont{Gauthier}} \bibnamefont{and}
  \bibinfo{author}{\bibfnamefont{M.A.}~\bibnamefont{Neifeld}},
  \bibinfo{journal}{Phys. Rev. Lett.} \textbf{\bibinfo{volume}{94}},
  \bibinfo{pages}{053902} (\bibinfo{year}{2005}).


\bibitem[{\citenamefont{Ranfagni et~al.}(2006)\citenamefont{Ranfagni et al}}]{ranfagni06}
\bibinfo{author}{\bibfnamefont{A.}~\bibnamefont{Ranfagni}},
  \bibinfo{author}{\bibfnamefont{P.}~\bibnamefont{Fabeni}},
  \bibinfo{author}{\bibfnamefont{G.P.}~\bibnamefont{Pazzi}},
  \bibinfo{author}{\bibfnamefont{A.M.}~\bibnamefont{Ricci}},
  \bibinfo{author}{\bibfnamefont{R.}~\bibnamefont{Trinci}},  
  \bibinfo{author}{\bibfnamefont{R.}~\bibnamefont{Mignani}},
  \bibinfo{author}{\bibfnamefont{R.}~\bibnamefont{Ruggeri}} \bibnamefont{and}
  \bibinfo{author}{\bibfnamefont{F.}~\bibnamefont{Cardone}},
  \bibinfo{journal}{Phys. Lett. A} \textbf{\bibinfo{volume}{352}},
  \bibinfo{pages}{473} (\bibinfo{year}{2006}).



\bibitem[{\citenamefont{Icsevgi et~al.}(1969)\citenamefont{Icsevgi and Lamb}}]{icsevgi}
\bibinfo{author}{\bibfnamefont{A.}~\bibnamefont{Icsevgi}} \bibnamefont{and}
  \bibinfo{author}{\bibfnamefont{W.E.}~\bibnamefont{Lamb}},
  \bibinfo{journal}{Phys. Rev.} \textbf{\bibinfo{volume}{185(2)}},
  \bibinfo{pages}{517} (\bibinfo{year}{1969}).


\bibitem[{\citenamefont{wang et~al.}(2000)\citenamefont{Wang, Kuzmich and Dogariu}}]{wang00}
\bibinfo{author}{\bibfnamefont{L.J.}~\bibnamefont{Wang}},
  \bibinfo{author}{\bibfnamefont{A.}~\bibnamefont{Kuzmich}} \bibnamefont{and}
  \bibinfo{author}{\bibfnamefont{A.}~\bibnamefont{Dogariu}},
  \bibinfo{journal}{Nature (London)} \textbf{\bibinfo{volume}{406}},
  \bibinfo{pages}{277} (\bibinfo{year}{2000}).

\bibitem[{\citenamefont{Janowicz et~al.}(2006)\citenamefont{Janowicz and Mostowski}}]{janowicz}
\bibinfo{author}{\bibfnamefont{M.}~\bibnamefont{Janowicz}} \bibnamefont{and}
  \bibinfo{author}{\bibfnamefont{J.}~\bibnamefont{Mostowski}},
  \bibinfo{journal}{Phys. Rev. E} \textbf{\bibinfo{volume}{73}},
  \bibinfo{pages}{046613} (\bibinfo{year}{2006}).

\bibitem[{\citenamefont{Huang et~al.}(2008)\citenamefont{Huang, Hang and Deng}}]{huang}
  \bibinfo{author}{\bibfnamefont{G.}~\bibnamefont{Huang}},
\bibinfo{author}{\bibfnamefont{Ch.}~\bibnamefont{Hang}} \bibnamefont{and}
  \bibinfo{author}{\bibfnamefont{L.}~\bibnamefont{Deng}},
  \bibinfo{journal}{Phys. Rev. A} \textbf{\bibinfo{volume}{77}},
  \bibinfo{pages}{011803(R)} (\bibinfo{year}{2008}).

\bibitem[{\citenamefont{solli et~al.}(2003)\citenamefont{Solli et al}}]{solli03}
\bibinfo{author}{\bibfnamefont{D.R.}~\bibnamefont{Solli}},
\bibinfo{author}{\bibfnamefont{C.F.}~\bibnamefont{McCormick}},
\bibinfo{author}{\bibfnamefont{C.}~\bibnamefont{Ropers}},
\bibinfo{author}{\bibfnamefont{J.J.}~\bibnamefont{Morehead}},
  \bibinfo{author}{\bibfnamefont{R.Y.}~\bibnamefont{Chiao}} \bibnamefont{and}
  \bibinfo{author}{\bibfnamefont{J.M.}~\bibnamefont{Hickmann}},
  \bibinfo{journal}{Phys. Rev. Lett.} \textbf{\bibinfo{volume}{91}},
  \bibinfo{pages}{143906} (\bibinfo{year}{2003}).

\bibitem[{\citenamefont{Brunner et~al.}(2004)\citenamefont{Brunner,Scarani, Wegm\"uller, Legr\'e and Gisin}}]{brunner}
\bibinfo{author}{\bibfnamefont{N.}~\bibnamefont{Brunner}},
  \bibinfo{author}{\bibfnamefont{V.}~\bibnamefont{Scarani}},
  \bibinfo{author}{\bibfnamefont{M.}~\bibnamefont{Wegm\"uller}},
  \bibinfo{author}{\bibfnamefont{M.}~\bibnamefont{Legr\'e}} \bibnamefont{and}
  \bibinfo{author}{\bibfnamefont{N.}~\bibnamefont{Gisin}},
  \bibinfo{journal}{Phys. Rev. Lett.} \textbf{\bibinfo{volume}{93}},
  \bibinfo{pages}{203902} (\bibinfo{year}{2004}).

\bibitem[{\citenamefont{halvorsen et~al.}(2008)\citenamefont{Halvorsen and Leinaas}}]{halvorsen}
  \bibinfo{author}{\bibfnamefont{T.G.}~\bibnamefont{Halvorsen}} \bibnamefont{and}
  \bibinfo{author}{\bibfnamefont{J.M.}~\bibnamefont{Leinaas}},
  \bibinfo{journal}{Phys. Rev. A} \textbf{\bibinfo{volume}{77}},
  \bibinfo{pages}{023808} (\bibinfo{year}{2008}).

\bibitem[{\citenamefont{Kulkarni et~al.}(2004)\citenamefont{Kulkarni et al}}]{kulkarni}
\bibinfo{author}{\bibfnamefont{M.}~\bibnamefont{Kulkarni}},
  \bibinfo{author}{\bibfnamefont{N.}~\bibnamefont{Seshadri}},
  \bibinfo{author}{\bibfnamefont{V.S.C.}~\bibnamefont{Manga Rao}} \bibnamefont{and}
  \bibinfo{author}{\bibfnamefont{S.}~\bibnamefont{Dutta Gupta}},
  \bibinfo{journal}{J. Mod. Opt.} \textbf{\bibinfo{volume}{51}},
  \bibinfo{pages}{549} (\bibinfo{year}{2004}).


\bibitem[{\citenamefont{safian et~al.}(2006)\citenamefont{Safian, Sarris, Mojahedi}}]{safian}
  \bibinfo{author}{\bibfnamefont{R.}~\bibnamefont{Safian}},
  \bibinfo{author}{\bibfnamefont{C.D.}~\bibnamefont{Sarris}} \bibnamefont{and}
  \bibinfo{author}{\bibfnamefont{M.}~\bibnamefont{Mojahedi}},
  \bibinfo{journal}{Phys. Rev. E} \textbf{\bibinfo{volume}{73}},
  \bibinfo{pages}{066602} (\bibinfo{year}{2006}).

\bibitem[{\citenamefont{Bigelow et~al.}(2006)\citenamefont{Bigelow, Lepeshkin, Shin and Boyd}}]{bigelow}
  \bibinfo{author}{\bibfnamefont{M.S.}~\bibnamefont{Bigelow}},
\bibinfo{author}{\bibfnamefont{N.N.}~\bibnamefont{Lepeshkin}}, 
\bibinfo{author}{\bibfnamefont{H.}~\bibnamefont{Shin}} \bibnamefont{and}
  \bibinfo{author}{\bibfnamefont{R.W.}~\bibnamefont{Boyd}},
  \bibinfo{journal}{J. Phys.: Condens. Matter} \textbf{\bibinfo{volume}{18}},
  \bibinfo{pages}{3117} (\bibinfo{year}{2006}).

\bibitem[{\citenamefont{Talukder et~al.}(2005)\citenamefont{Talukder, Haruta, Tomita}}]{talukder}
  \bibinfo{author}{\bibfnamefont{A.I.}~\bibnamefont{Talukder}},
  \bibinfo{author}{\bibfnamefont{T.}~\bibnamefont{Haruta}} \bibnamefont{and}
  \bibinfo{author}{\bibfnamefont{M.}~\bibnamefont{Tomita}},
  \bibinfo{journal}{Phys. Rev. Lett.} \textbf{\bibinfo{volume}{94}},
  \bibinfo{pages}{223901} (\bibinfo{year}{2005}).

\bibitem[{\citenamefont{Mugnai et~al.}(2000)\citenamefont{Mugnai, Ranfagni, Ruggeri}}]{mugnai00}
  \bibinfo{author}{\bibfnamefont{D.}~\bibnamefont{Mugnai}},
\bibinfo{author}{\bibfnamefont{A.}~\bibnamefont{Ranfagni}} \bibnamefont{and}
  \bibinfo{author}{\bibfnamefont{R.}~\bibnamefont{Ruggeri}},
  \bibinfo{journal}{Phys. Rev. Lett.} \textbf{\bibinfo{volume}{84}},
  \bibinfo{pages}{4830} (\bibinfo{year}{2000}).

\bibitem[{\citenamefont{Mugnai et~al.}(2005)\citenamefont{Mugnai and Mochi}}]{mugnai05}
\bibinfo{author}{\bibfnamefont{D.}~\bibnamefont{Mugnai}} \bibnamefont{and}
  \bibinfo{author}{\bibfnamefont{I.}~\bibnamefont{Mochi}},
  \bibinfo{journal}{Phys. Rev. E} \textbf{\bibinfo{volume}{73}},
  \bibinfo{pages}{016606} (\bibinfo{year}{2006}).

\bibitem[{\citenamefont{Carey et~al.}(2000)\citenamefont{Carey, Zawadzka, Jaroszynski, and Wynne}}]{carey00}
\bibinfo{author}{\bibfnamefont{J.~J.} \bibnamefont{Carey}},
  \bibinfo{author}{\bibfnamefont{J.}~\bibnamefont{Zawadzka}},
  \bibinfo{author}{\bibfnamefont{D.~A.} \bibnamefont{Jaroszynski}}
  \bibnamefont{and} \bibinfo{author}{\bibfnamefont{K.}~\bibnamefont{Wynne}},
  \bibinfo{journal}{Phys. Rev. Lett.} \textbf{\bibinfo{volume}{84}},
  \bibinfo{pages}{1431} (\bibinfo{year}{2000}).

\bibitem[{\citenamefont{Mochan and Brudny}(2001)}]{mochan01}
\bibinfo{author}{\bibnamefont{W.L.}~\bibnamefont{Moch\'an}} \bibnamefont{and}
  \bibinfo{author}{\bibnamefont{V.L.}~\bibnamefont{Brudny}}, 
  \bibinfo{journal}{Phys.Rev.Lett} \textbf{\bibinfo{volume}{87}},
  \bibinfo{pages}{119101} (\bibinfo{year}{2001}).

\bibitem[{\citenamefont{Carey et~al.}(2001)\citenamefont{Carey, Zawadzka,  Jaroszynski and Wynne}}]{carey01}
\bibinfo{author}{\bibfnamefont{J.~J.} \bibnamefont{Carey}},
  \bibinfo{author}{\bibfnamefont{J.}~\bibnamefont{Zawadzka}},
  \bibinfo{author}{\bibfnamefont{D.~A.} \bibnamefont{Jaroszynski}}
  \bibnamefont{and} \bibinfo{author}{\bibfnamefont{K.}~\bibnamefont{Wynne}},
  \bibinfo{journal}{Phys. Rev. Lett.} \textbf{\bibinfo{volume}{87}},
  \bibinfo{pages}{119102} (\bibinfo{year}{2001}).

\bibitem[{\citenamefont{Shaarawi et~al.}(2002)\citenamefont{Shaarawi, Tawfik and Besieris}}]{shaarawi}
  \bibinfo{author}{\bibfnamefont{A. M.}~\bibnamefont{Shaarawi}},
\bibinfo{author}{\bibfnamefont{B. H.}~\bibnamefont{Tawfik}} \bibnamefont{and}
  \bibinfo{author}{\bibfnamefont{I. M.}~\bibnamefont{Besieris}},
  \bibinfo{journal}{Phys. Rev. E.} \textbf{\bibinfo{volume}{66}},
  \bibinfo{pages}{046626} (\bibinfo{year}{2002}).

\bibitem[{\citenamefont{Brudny and Mochan}(2001)}]{oe}
\bibinfo{author}{\bibnamefont{V.L.}~\bibnamefont{Brudny}} \bibnamefont{and}
  \bibinfo{author}{\bibnamefont{W.L.}~\bibnamefont{Moch\'an}}, 
  \bibinfo{journal}{Optics Express} \textbf{\bibinfo{volume}{19(11)}} 
  \bibinfo{pages}{561}(\bibinfo{year}{2001}).

\bibitem[{\citenamefont{Barbero et~al.}(2000)\citenamefont{Barbero et al}}]{barbero}
  \bibinfo{author}{\bibfnamefont{A. P.}~\bibnamefont{Barbero}}
  \bibinfo{author}{\bibfnamefont{H. E.}~\bibnamefont{Hern\'andez-Figueroa}} \bibnamefont{and}
  \bibinfo{author}{\bibfnamefont{E.}~\bibnamefont{Recami}},
  \bibinfo{journal}{Phys. Rev. E} \textbf{\bibinfo{volume}{62(6)}},
  \bibinfo{pages}{8628} (\bibinfo{year}{2000}).

\bibitem[{\citenamefont{Reiten et~al.}(2001)\citenamefont{Reiten et al}}]{reiten}
  \bibinfo{author}{\bibfnamefont{M. T.}~\bibnamefont{Reiten}}
  \bibinfo{author}{\bibfnamefont{D.}~\bibnamefont{Grischkowsky}} \bibnamefont{and}
  \bibinfo{author}{\bibfnamefont{R. A.}~\bibnamefont{Cheville}},
  \bibinfo{journal}{Phys. Rev. E} \textbf{\bibinfo{volume}{64}},
  \bibinfo{pages}{036604} (\bibinfo{year}{2001)}.

\bibitem[{\citenamefont{Balcou et~al.} (1997)\citenamefont{Balcou and Dutriaux}}]{balcou97}
  \bibinfo{author}{\bibfnamefont{Ph.}~\bibnamefont{Balcou}} \bibnamefont{and}
  \bibinfo{author}{\bibfnamefont{L.}~\bibnamefont{Dutriaux}},
  \bibinfo{journal}{Phys. Rev. Lett.} \textbf{\bibinfo{volume}{78(5)}},
  \bibinfo{pages}{851} (\bibinfo{year}{1997}).

\bibitem[{\citenamefont{Resch et~al.}(2001)\citenamefont{Resch, Lundeen and Steinberg}}]{resch}
  \bibinfo{author}{\bibfnamefont{K.J.}~\bibnamefont{Resch}},
\bibinfo{author}{\bibfnamefont{J.S.}~\bibnamefont{Lundeen}} \bibnamefont{and}
  \bibinfo{author}{\bibfnamefont{A.M.}~\bibnamefont{Steinberg}},
  \bibinfo{journal}{IEEE J. Quantum Electron.} \textbf{\bibinfo{volume}{37(6)}},
  \bibinfo{pages}{794} (\bibinfo{year}{2001}).



\bibitem[{\citenamefont{Morse and Feshbach}(alguno)}]{inmorse}
\bibinfo{author}{\bibnamefont{Phillip M.}~\bibnamefont{Morse}} \bibnamefont{and}
  \bibinfo{author}{\bibnamefont{Herman}~\bibnamefont{Feshbach}}, \emph{\bibinfo{title}{Methods of Theoretical Physics}} 
(\bibinfo{publisher}{Mc.Graw-Hill},
  \bibinfo{address}{New York}, \bibinfo{year}{1953}).



\end{thebibliography}

\end{document}